\documentclass[jap, preprint]{revtex4-1}

\begin{document}

\title{Surface Effects on the Mechanical Elongation of AuCu Nanowires: De-alloying and the Formation of Mixed Suspended Atomic Chains }

\author{M. J. Lagos$^{1,2,+}$}
\author{P. A. S. Autreto$^{1}$}
\author{J. Bettini$^{2}$}
\author{F. Sato$^{3}$}
\author{S. O. Dantas$^{3}$}
\author{D. S. Galvao$^{1}$}
\author{D. Ugarte$^{1}$}
\email{galvao@ifi.unicamp.br}

\affiliation{$^{1}$Instituto de F\'{\i}sica Gleb Wataghin, Universidade Estadual de Campinas, R. Sergio B. de Holanda 777, 13083-859 Campinas-SP, Brazil}

\affiliation{$^{2}$Laborat\'{o}rio Nacional de
Nanotecnologia-LNNANO, 13083-970 Campinas-SP, Brazil}

\affiliation{$^{3}$Departamento de F\'{\i}sica, ICE, Universidade
Federal de Juiz de Fora, 36036-330 Juiz de Fora-MG, Brazil}

\affiliation{$^{+}$Present Address: Institute for Advanced
Materials, Devices and Nanotechnolgy- IAMND, Rutgers University,
Piscataway NJ 00854, USA}

\date{\today}

\begin{abstract}
We report here an atomistic study of the mechanical deformation of
Au$_{x}$Cu$_{(1-x)}$  atomic-size wires (NWs) by means of high resolution
transmission electron microscopy (HRTEM) experiments. Molecular dynamics
simulations were also carried out in order to obtain deeper insights on the
dynamical properties of stretched NWs. The mechanical properties are
significantly dependent on the chemical composition that evolves in time
at the junction; some structures exhibit a remarkable de-alloying
behavior. Also, our results represent the first experimental realization of mixed
linear atomic chains (LACs) among transition and noble metals; in particular,  surface energies induce
chemical gradients on NW surfaces that can be exploited to control the
relative LAC compositions (different number of gold and copper atoms). The
implications of these results for nanocatalysis and spin transport of
one-atom-thick metal wires are addressed.

\end{abstract}

\maketitle

\section{Introduction}

Predicting the mechanical behavior of a strained nanoscale volume of matter is essential for many nanotechnological
applications \cite{nmnni}. This has stimulated an intense  study
of mechanical elongation of atomic-size metal nanowires (NWs) \cite{reviewNW}.
In this range size ($\sim$1-2 nm in diameter), surface energy plays a dominant role and, factors that are neglected in macroscopic theory,
such as size and shape, determine deformation mechanisms.
For example, surface energy can induce strengthening and asymmetrical mechanical response \cite{varleiprl,greer,PRLdefect,ActMat}.
In addition, the high surface/volume ratio (SVR) may lead to the generation of
anomalous helicoidal or tubular nanostructures during deformation \cite{scientaka, nattube}. 

Alloying or doping are routinely utilized to improve the mechanical
resistance of metals (solute strengthening) \cite{callister}. However, these
manipulations are very difficult to apply to nanosystems due to the huge SVR, which may
promote composition gradients, or even the expelling of impurities 
\cite{marksnl,natnanoAuAg,natpurif}. In addition, most of our knowledge on metal alloy nanosystem is associated
with heterogeneous catalysts, where the use of alloy nanoparticles (NPs)
represents an active research field. Nanoscale mechanical deformation of
alloys represents a quite complex topic, as the constant injection of
elastic energy into the system may be relaxed through a wide variety of 
structural, physical  and chemical mechanisms. We also must keep in mind that the analysis of compositional gradients and segregation in alloy metal nanoparticles  in heterogeneous catalysis still represent a question that awaits for  a reliable answer. A recent cutting edge study reports the analyses of composition gradients in metal nanoparticles exploiting X ray TEM tomography  \cite{slater2014}. A recent cutting edge study reports the analyses of composition gradients in metal nanoparticles exploiting X ray TEM tomography \cite{slater2014}.
 In this way, this research remains
quite challenging, as we must analyze the complex interplay between
elastic, electronic and surface energy contributions. 

Here, we present a detailed study of atomic structure
evolution of Au-Cu alloy NWs following tensile
deformation by means of high resolution
electron microscopy (HRTEM). Molecular dynamics simulations were also carried out in order to
analyze the dynamics of atomistic processes involved in the nanoalloy
physical and chemical modifications.

\section{Methodology}

We have generated metal NWs from alloy bimetallic
films (Au$_x$Cu$_{(1-x)}$ (0 $<$ x $<$ 1)) following
the experimental procedure introduced by Takayanagi's group
\cite{PRLtaka}. 
Initially, holes are opened at several points in a self-supported metal ﬁlm by focusing the microscope electron beam ($~300 A/cm^{2}$); in this manner, nanometric constrictions (bridges) are formed between them. Then, the microscope beam current density is reduced to standard operation values (10–30 A/cm2) for image acquisition; in this range of beam current density, the HRTEM sample temperature is estimated to be within $300-350$ K \cite{carter} .  The spontaneous elongation and rupture of the nanowires is acquired using a high-sensitive TV camera (Gatan 622SC, 30 frames/s) and a standard video/DVD recorder. It is important to emphasize that this experimental procedure allows the acquisition of time-resolved atomic-resolution-imaging of NWs with a remarkable quality; nevertheless, it is not possible neither to measure the force being applied nor to control the stretching direction  \cite{rodriguesbook}. Usually the NW stretching and fracture occurs with average displacement rates of 0.1-1 nm per second. The worked described here used bimetallic $Au_{x}Cu_{1-x}$ alloy thin films as initial sample to generate NWs in situ in the HRTEM (JEM-3010 URP 300 kV, 0.17 nm point resolution).

Polycrystalline $Au_{x}Cu_{1-x}$ ﬁlms ($30-50$ nm in thickness) have been prepared by thermal co-evaporation in a standard vacuum evaporator ($10ˆ{-7}$ mbar). A quartz crystal monitor was used to set the evaporation rate of each individual metal source and, subsequently, to measure the equivalent thickness of the ﬁlm. Owing to higher cooling rates associated with the film deposition the bimetallic films are expected to consist of a solid solution with random distribution of gold and copper atoms \cite{porterbook}, what was confirmed by micro-electron diffraction results. To prevent possible oxidation by exposure to ambient conditions, the bi-metallic ﬁlms were sandwiched between two (3-nm-thick) amorphous carbon thin layers. Before generating the NWs, the carbon layers are removed by strong electron irradiation \cite{prbCuchain} inside the HRTEM. 
The structural characterization has been performed by means micro-electron diffraction (JEM 2100 ARP, operated at 200 kV). In our experiments, the electron diffraction patterns (DP) were acquired from a region of ~800 nm in diameter and recorded using a CCD camera (Gatan ES500W) (more details see Figure S1 in the Suplementary Material).
We have also measured the chemical composition of synthesized alloy films using Energy-Dispersive X-ray Spectroscopy (EDS); the $Au_{x}Cu_{1-x}$ alloy films were supported over conventional molybdenum TEM grids to avoid spurious x-ray signal. In particular, we analyzed several localized regions inside the initial illuminated area used for electron diffraction studies and, the observed atomic composition variations were within the typical composition error bar ($~5\%$, using Cliff-Lorimer method without absorption correction \cite{carter}). Also, the measured compositions were in very good agreement with electron diffraction estimation using Vegard’s law. However, we have observed significant composition changes when comparing measurements performed in pristine alloy thin films and, after the in-situ formation of NW’s (i.e. after intense electron beam irradiation during several hours). It is important to highlight that the electron irradiation necessary to prepare the metal alloy film for a NW study requires a several hours long electron beam irradiation. In contrast, the final NW elongation and rupture processes recorded by the experimental videos last, at most, $3-5$ minutes at a much lower electron beam intensity (see description of electron microscopy works in Section 1.a of Supplementary Materials); then, it is reasonable to think that no significant chemical composition change occurs during the nanowires imaging study. In this sense, we have assumed that the final EDS estimated concentration is a good value to describe the NW composition, and it has been used to describe the nanowires.

We have also carried out molecular dynamics simulations to
gather deeper insights on the atomistic
processes occurring during the alloy NW elongation. A
tight-binding molecular dynamics methodology based on the
second-moment approximation (TB-SMA) \cite{CR,TAB} was used to
analyze the elongation structural evolution.  The
theoretical methodology has already been described in detail by Sato \textit{et al.}
\cite{sato2005}; this approach has
proved to be very effective for the study of Au and Cu NWs
\cite{coura2004,JuanCu, prbCuchain}.

\section{Results}

The in-situ HRTEM experiments indicate that Au$_{x}$Cu$_{(1-x)}$
alloy NWs can deform along only three [111], [100] and [110]
crystallographic directions.  NWs elongated along [111] and [100]
directions generate by-pyramidal constrictions that always evolve
into an atomic contact or linear atomic chains (LACs); this behavior is identical to
pure Au or Cu wires \cite{varleiprl,JuanCu}. In contrast, alloy
NWs stretched along the [110] direction (hereafter noted as [110]
NWs) display a concentration dependent structural behavior. While
[110] Au NWs display rod-like morphology and break abruptly when
formed by 3-4 atomic layer thick \cite{varleiprl,coura2004},
 alloy NWs (both Au$_{0.55}$Cu$_{0.45}$ and Au$_{0.2}$Cu$_{0.8}$)
 display a structural sequence typical of pure Cu wires
\cite{JuanCu,prbCuchain}: 1) rod-like wire ; 2) by-pyramidal
constrictions and, 3) a final one-atom-thick contact or LAC
formation (Figures 1b,c, and corresponding videos in the
SM). This indicates that a Cu
content of $\sim$ 45$\%$ is enough to trigger typical copper
behavior, consequently modifying the rupture mode from brittle to
ductile. These results are quite different from the Au$_{x}$Ag$_{(1-x)}$ NW case, where
a much higher Ag content ($\sim$ 80$\%$) was necessary to
reveal silver nanowire characteristics \cite{natnanoAuAg}.

From a mechanical point of view, it
is important to analyze the active deformation mechanisms
of atomic-size alloy wires.
Concerning bulk material, Au and Cu are Face Centered Cubic (FCC) metals and, plastic deformation
occurs mostly by the gliding of compact (111) atomic planes along
[112]-type directions. In particular, partial edge dislocations
(PDs) are formed, which encapsulate a stacking fault (SF) ribbon.
In tiny gold nanorods, where diameter ($\sim$1 nm) is smaller than the SF ribbon width,
((\emph{d}) $\sim$2-3 nm in bulk \cite{callister}), plastic deformation occurs by
the formation of planar defects that generate a compact
glide (block on a block) of the (111) planes by (1/6)[112] \cite{PRLdefect,landman}.
In these very tiny wires thermal energy at room temperature is enough to recombine these planar faults
\cite{ovidko,PRLdefect} and, pure Au and pure Cu NWs
stay defect free when stretched at 300 K \cite{varleiprl,JuanCu}.
Concerning macroscopic alloys, it is well
known that alloying influences drastically, the elastic modulus
and yield strength \cite{callister}. On this basis we could expect
that, in Au-Cu alloy NWs, energy barrier blocking planar defects
should be higher. In fact, we have observed
the formation of planar defects at room temperature in some NWs
with Au$_{0.55}$Cu$_{0.45}$ composition (Figure
2a,b). Nevertheless, many alloy NWs also displayed
defect free structures; this may be associated
with subtle local variations of chemical composition inside the alloy thin
film or, be even induced during the wire elongation \cite{natnanoAuAg}.
Figures 2c and 2d
show some interesting images of Au$_{0.20}$Cu$_{0.80}$ NWs. Note
several darker dots at the NW apexes. This might be associated with
the formation of several small gold clusters during the NW
elongation (gold atoms are expected to generate darker dots in the images).
Accordingly, our video recordings
show that these clusters move
slowly during the mechanical elongation, suggesting that they may
be located on the NW surface (see Supplementary Material (SM)).
In fact, several theoretical studies of Au-Cu nanoparticles
have predicted the migration of gold atoms to the surface in order minimize surface energy
\cite{marksnl,PRLyacaman}. Certainly, the gold lower surface energy
\cite{artigosurfenergy} and lower diffusion barrier drive gold atoms
migration to the wire surface during the mechanical deformation.

Molecular dynamics simulations can provide additional insight into atom
reorganization and redistribution during alloy NW elongation.
Figure 3a shows a sequence of
snapshots of the stretching of Au$_{0.5}$Cu$_{0.5}$
NW along [110] direction. Initially, a by-pyramidal
constrictions is formed, in good agreement with experimental
observations; then, a long NW is generated. It is important to
emphasize that most of the NW gold atoms are located on the NW
surface, the inset shows a cross-sectional view where it is
clear that Au atoms enclose a chain of Cu atoms. Finally, a
one-atom-thick contact is formed before breaking. During the
elongation, a clear gold enrichment of the narrowest wire regions
can be measured  (see also data in the SM). Figure 3b shows a similar Au surface migration
effect in Au$_{0.2}$Cu$_{0.8}$ NW stretched along [110] axis.
However, the behavior changes, because the initial Au content is rather low (20$\%$) and,
 there are not enough available
gold atoms to cover the whole surface \cite{marksnl}.
Finally, Figure 3c illustrates an Au$_{0.2}$Cu$_{0.8}$ NW
being elongated along the [100] direction, which becomes gradually
thinner until forming an atomic contact. The formation of small
gold clusters (3-5 atoms) on the NW surface can be clearly observed.
The segregated gold clusters remain coalesced and sometimes diffuse
slowly on the NW surface. Briefly, the simulations
have revealed two effects associated with
 local composition changes: (i) surface segregation of Au
atoms and, (ii) gold clustering. These effects are in excellent agreement with
the experimental observations displayed in Figures 1 and 2. However,
we must keep in mind that electron beam induced effects that may also
influence atom diffusion or segregation. We think that these
effects are negligible in our experiments due to the rather short
duration of the experiments (less than 1 minute for the elongation
and rupture of the 1-nm-wide alloy wires, see a detailed discussion
in the SM)

From a more fundamental point of view, the formation of one-atom-thick nanowires  
containing different metal atomic species, represent one of the most 
interesting nanosystems to
study 1-D quantum physics. So far, mixed suspended linear atom chains (LACs) were experimentally produced
only with gold and silver \cite{natnanoAuAg}. 
Mixed LACs containing gold and transition metal atoms open the possibility to address 
excellent physical and chemical questions such as nanomagnetism, spin transport, s-d bonding in low dimensional systems, etc. 
From this perspective, the one-atom thick wire generated from the Au-Cu alloy nanowires observed in our study 
represent an excellent case study.
Our HRTEM results indicate
that suspended chains display variations in the intensity/contrast
at atomic positions (see Figure 1c and SM), suggesting that they 
should be formed by both gold and copper atoms.  A quantitative comparison between experimental and
simulated HRTEM image intensities confirms this interpretation.  The interatomic distances in the LAC
are in the [0.25-0.32] nm range, what agrees with
impurity-free gold and copper chains.
\cite{JuanCu,prlAuchains,nanotechAuchains}.
Light possible impurity atoms such as C,O,N should produce a much lower contrast than
the signal noise ratio experimentally observed (see details in SM).

A natural question arises, can we control the
LAC chemical composition by selecting the proper alloy, wire shape, elongation direction, etc.? 
Previous theoretical studies have shown that most of the atoms
composing suspended chains come from the outermost layers
for Au nanowires \cite{LACatoms}.
With this idea in mind, we have analyzed our theoretical simulations and looked at the chemical composition (pure Au or Cu, or alloyed chains)
of LACs generated along different stretching directions for different alloy compositions
(Au$_{0.5}$Cu$_{0.5}$ and Au$_{0.2}$Cu$_{0.8}$).
Most of suspended atomic chains were composed by two hanging
atoms, while seldom rather long chains (4-5 atoms) were observed.
Although, the rather low available statistics, a clear alloy
composition and elongation direction dependence
shows up (see results in Table 1).
LACs generated from [100] NWs show the tendency to be either pure Au or pure Cu
depending on the alloy mixture.
In contrast, mixed LACs dominate the occurrence along [111] axis
for both studied alloys. Finally, [110] alloy nanowires show a slight tendency
to produce pure Au chains, followed by alloyed chains.

In order to understand these simulation results, we must first consider that alloy composition
influences NW morphologies through changes in
the surface energy of the different crystallographic facets \cite{marksnl}.
A gold particle should be a cuboctahedra with regular hexagonal facets
(minimal energy planes are \{111\})\cite{marksnl}, schema at left in Figure 4a)),
while Cu nanoparticle \cite{urban} have a morphology dominated by
 \{100\} facets  (center and right octahedra in Figure 4a).
In first approximation, an alloy nanosystem behavior must be somewhat in between these two extrema \cite{marksnl}.
In addition, as diffusion and migration is enhanced in nanosystems,
it is reasonable to make the hypothesis that the wires will have a spontaneous tendency to have composition gradient in the volume and on the surfaces.
In particular, we can assume that \{hkl\}  facets will accommodate more atoms of the chemical species that minimize the facets surface energy,
what would lead to Au (Cu) rich  \{111\} (\{100\}) facets in an Au-Cu alloy NW (or NP).
A gold-rich-alloy NW along [100] should have a pyramidal shape of
 square base defined by four triangular  \{111\} facets (see left side in Figure 4b \cite{varleiprl}).
 These facets will become Au rich during elongation, what finally it
will enhance the formation of pure Au LACs.
In contrast, a Cu rich [100] NW should display a rod-like shape with
a square cross section, the surface being formed by four \{100\} facet (see right side in Figure 4b),
then having a tendency to generate mostly pure Cu LACS.
Wires along [110] direction
will have a hexagonal cross section formed by both \{111\} and \{100\}
facets, with a relative weight that varies from Au to Cu; this explain why pure Au and mixed chains  are formed along [110] elongation axis.
Finally, alloy NWs formed along [111] will generate at some moment a triangular cross section (see for example the \{111\} facet shape of octahedra at the center of image 4a). This triangular facet is surrounded by 3 \{100\} (Cu rich) facets and the tip of 3 \{111\}(Au rich) facets,
what can provide both Au and Cu atoms. This kind of wire must be expected to generate mostly mixed LAC's, as in fact, we observed in the simulations.

In summary, we have observed that alloy Au$_{x}$Cu$_{(1-x)}$ NWs show a strong
concentration dependence mechanical behavior. Approximately a $\sim$ 45$\%$ Cu content is required
to trigger copper-alike mechanical behavior.
For the tiny alloy NWs studied here ($\sim$nm in diameter) surface energy contribution is so important that can induce gold enrichment and, even gold surface segregation
during elongation. The formation of suspended atom chains containing Au and Cu atoms was experimentally revealed.
Molecular dynamics simulations suggest that it is possible to control the LAC chemical composition 
by choosing the appropriate alloy composition and NW elongation direction.
This would exploit the spontaneous
formation of chemical composition gradients (or preferential chemical enrichment of each family of crystallographic facets) on the NW surface. 
This phenomena can certainly be expected to also happen in alloy nanoparticles and, may modify significatively the reactivity  and/or catalytic activity in heterogeneous catalysis.
From another point of view, the possibility to generate in a reasonable controlled way alloy LACs containing Au and magnetic transition metals may open new opportunities for 
the study scattering and spin transport in one-atom-thick metal wires.

P.C. Silva is acknowledged for assistance during HRTEM and sample
preparation work.  We acknowledge financial support from LNLS, FAPESP and CNPq.
The authors thank the Center for Computational Engineering and Sciences
at Unicamp for financial support through the FAPESP/CEPID Grant 2013/08293-7.

{}

\newpage
\clearpage

\begin{table}
\caption{Statistical analysis of LAC formation from the molecular
dynamics simulations. The three numbers (X/Y/Z) indicate the
number of LAC formed and composed only of
gold, mix Au/Cu or copper, respectively.} \label{tab1}
\begin{tabular}{cccc}
\hline
Au$_{x}$ Cu$_{y}$   &   [100]  &   [110]  &  [111]  \\
\hline
       50/50      & 6/3/0 & 6/4/0 & 2/6/2 \\
       80/20      & 2/2/7 & 4/3/2 & 3/7/1 \\
\hline
\end{tabular}
\end{table}

\begin{figure}
\caption{Sequence of atomic resolution images associated with the
elongation and thinning of rod-like [110] Au-Cu NWs as a function
of chemical composition (atom positions appear dark).
(a) Au$_{0.55}$Cu$_{0.45}$ and (b)
Au$_{0.2}$Cu$_{0.8}$ NWs form by-pyramidal constrictions and
evolve into either suspended atomic chains or atomic contact.
(c) closer view of suspended linear atomic chain generated from a
Au$_{0.55}$Cu$_{0.45}$ NW; we can see that
the atom at the left of the LAC  (arrowed) display a different contrast
from the other suspended atoms suggesting the formation of mixed Au-Cu chains.
}
\end{figure}

\begin{figure}
\caption{Typical HRTEM images of Au$_{0.55}$Cu$_{0.45}$ NW's
displaying planar defects (a) and twins (b). Defects are indicated
by black arrows and, they can also be identified by a
discontinuity in the atomic planes. (c,d) HRTEM images of stretched
Au$_{0.2}$Cu$_{0.8}$ NWs which indicate the formation of small
gold clusters in the NW surface shown inside the ellipses.
Atom positions appear dark.
}
\end{figure}

\begin{figure}
\caption{ Sequence of snapshots associated with the theoretical
simulations of bi-metallic Au-Cu nanowire elongation. Gold and copper atoms are
represented by yellow and dark gray balls, respectively.
Figures (a) and (b) correspond to the elongation  along [110] axis of Au$_{0.55}$Cu$_{0.45}$
and Au$_{0.2}$Cu${0.8}$ NWs, respectively. Note that in both cases
gold atoms have a tendency to occupy the NW surface. Figure (c) illustrates the
formation of small gold clusters (3-5 atoms) on the NW surface
(indicated by arrows) during the elongation of a Au$_{0.2}$Cu${0.8}$ NW along [100] axis.
The corresponding animations can be found in the SM}
\end{figure}

\newpage
\clearpage

\begin{figure}
\caption{ a) From left to right, shapes of cuboctahedral FCC nanoparticles
increasing the relevance of \{100\} facets in relation to \{111\} facets. b) Scheme of the
possible [100] alloy NW; a pyramidal shape  containing only \{111\} surfaces, expected for a gold rich wire and,
a rod-like wire of square cross-section expected for Cu rich NW (only \{100\} facets cover the surface).
(c) Qualitative hexagonal cross-section for a pillar-like wire expected along [110] direction,
note that this rod is formed by both \{111\} and \{100\} facets (see text for discussions).  }
\end{figure}


\begin{thebibliography}{}

\bibitem{nmnni} C. Alloca and D. Smith, \textit{Instrumentation and Metrology for
Nanotechnology, Report of the National Nanotechnology Initiative}
(available from www.nano.gov, 2005), Ch. 3.

\bibitem{reviewNW} G. Rubio, N. Agra\"{i}t, and S. Vieira,
\textit{Phys. Rev. Lett.} \textbf{76}, 2302 (1996).

\bibitem{varleiprl} V. Rodrigues, T. F\"{u}hrer and D. Ugarte,
\textit{Phys. Rev. Lett.} \textbf{85}, 4124 (2000).

\bibitem{greer} S. Brinckmann, J.-Y. Kim and, J. R. Greer,
\textit{Phys. Rev. Lett. } \textbf{100}, 155502 (2008).

\bibitem{PRLdefect} M. J. Lagos, F. Sato, D. S. Galv\~{a}o and D. Ugarte,
\textit{Phys. Rev. Lett.} \textbf{106}, 055501 (2011).

\bibitem{ActMat} K Sieradzki, A . Rinaldi, C. Friesen and P. Peralta, \textit{Acta Mater.} \textbf{54}, 4533 (2006).

\bibitem{scientaka} Y. Kondo, and K. Takayanagi, \textit{Science} \textbf{289}, 606 (2000).

\bibitem{nattube} M. J. Lagos, J. Bettini, F. Sato, V. Rodrigues, D. S. Galv\~{a}o and D. Ugarte,
\textit{Nature Nanotech.} \textbf{4}, 149 (2009).

\bibitem{callister}W. D. Callister,  \textit{Materials science and engineering : an introduction}
(J. Wiley, New York, 2003).

\bibitem{marksnl} E. Ringe, R.P Van Duyne and L. D. Marks,
\textit{Nano Lett.} \textbf{11}, 3399 (2012).

\bibitem{natnanoAuAg}  J. Bettini, F. Sato, P. Z. Coura, S. O. Dantas, D. S. Galv\~{a}o and D. Ugarte,
\textit{Nature Nanotech.} \textbf{1}, 182 (2006).

\bibitem{natpurif}  S. C. Erwin, L. Zu, M. I. Haftel, A. L. Efros, T. A. Kennedy and D. J. Norris,
\textit{Nature} \textbf{436}, 91 (2005).

\bibitem{slater2014}  T. J. A. Slater, A. Macedo, S. L. M. Schroeder, M. Grace Burke, P. OBrien, P. H. C. Camargo and S. J. Haigh,
\textit{NanoLett.} \textbf{14}, 1921 (2014).

\bibitem{prbenomoto} A. Enomoto, S. Kurokawa, A. Sakai,
\textit{Phys. Rev. B} \textbf{65}, 125410 (2002).

\bibitem{PRLtaka} Y. Kondo, K. Takayanagi,
\textit{Phys. Rev. Lett.} \textbf{79}, 3455 (1997).

\bibitem{carter}C. B. Carter and, D.B. Williams  \textit{Transmission Eletron Microscopy}
(Springer, New York, 2009).

\bibitem{rodriguesbook} V. Rodrigues and, D. Ugarte  \textit{Nanowires and Nanobelts},
edited by Z. L. Wang ( Kluwer Acad. Publ., Boston, 2003), Vol. 1, Ch. 6.

\bibitem{porterbook} D. A. Porter and K. E. Easterling \textit{Phase Transformations in Metals and Alloys},
( Chapman and Hall, London, 1992).

\bibitem{prbCuchain} F. Sato, A. S. Moreira, J. Bettini, P. Z. Coura, S.O. Dantas, D. Ugarte and D.S. Galv\~{a}o,
\textit{Phys. Rev. B} \textbf{74}, 193401 (2006).

\bibitem{CR} F. Cleri and V. Rosato, Phys. Rev. B
\textbf{48}, 22 (1993).

\bibitem{TAB} D. Tom\`{a}nek, A. A. Aligia and C. A. Balseiro,
\textit{Phys. Rev. B} \textbf{32}, 5051 (1985).

\bibitem{sato2005} F. Sato, A. S. Moreira, P.Z. Coura, S.O. Dantas, S. B. Legoas, D. Ugarte and D.S. Galv\~{a}o:,
\textit{Appl. Phys A}  \textbf{81}, 1527 (2005).

\bibitem{coura2004} P.Z. Coura, S.B. Legoas, A.S. Moreira, F. Sato, V. Rodrigues, S.O. Dantas, D.Ugarte, and D.S. Galv\~{a}o,
\textit{Nano Lett.} \textbf{47}, 1187 (2004).

\bibitem{JuanCu} J.C. Gonz\'{a}lez, V. Rodrigues, J. Bettini, L.G.C. Rego, A.R. Rocha, P.Z. Coura, S.O. Dantas, F. Sato, D.S. Galv\~{a}o and D. Ugarte,
\textit{Phys. Rev. Lett.} \textbf{93}, 126103 (2004).



\bibitem{landman} U. Landman, W. D. Luedtke, N. A. Burnham, and R. J. Colton,
\textit{Science} \textbf{248}, 454 (1990).

\bibitem{ovidko} I. A. Ovid'ko and A. G. Sheinerman,
\textit{Reviews on Advanced Materials Science} \textbf{27}, 189 (2011).

\bibitem{rice} J. R. Rice, \textit{J. Mech. Phys. Solids} \textbf{40}, 239 (1992).

\bibitem{tadmor} E. B. Tadmor, S. Hai, \textit{J. Mech. Phys. Solids} \textbf{51}, 765 (2003).


\bibitem{PRLyacaman} J. L. Rodriguez-Lopez,  J. M. Montejano-Carrizales, U. Pal,J. F. Sanchez-Ramirez, H. E. Troiani, D. Garcia, M. Miki-Yoshida,
M. Jose-Yacaman,
\textit{Phys. Rev.Lett.}\textbf{92}, 196102
(2004).

\bibitem{artigosurfenergy} L. Vitos, A. V. Ruban, H. L. Skriver, J. Kollar,
 \textit{Surface Science} \textbf{411}, 186 (1998).

\bibitem{nanotechAuchains} M. J. Lagos, F. Sato, P. A. S. Autreto, D. S. Galv\~{a}o, V. Rodrigues, D. Ugarte,
\textit{Nanotechnology} \textbf{22}, 095705 (2011).

\bibitem{prlAuchains} S. B. Legoas, D. S. Galv\~{a}o, V. Rodrigues, D. Ugarte,
\textit{Phys. Rev. Lett.} \textbf{88}, 076105 (2002).

\bibitem{urban}
I. Lisiecki, A. Filankembo, H. Sack-Kongehl, K. Weiss, M. P.
Pileni and J. Urban, Phys. Rev. B \textbf{61}, 4968 (2000).

\bibitem{LACatoms} F. Sato, A. S. Moreira, P. Z. Coura, S.O. Dantas, S. B. Legoas, D. Ugarte and D.S. Galv\~{a}o,
\textit{Appl. Phys A} \textbf{81}, 1527 (2005).

\end{thebibliography}
\end{document}